\documentstyle[preprint,aps,epsf,floats]{revtex}  

\newcommand{\nn}{\nonumber}

\begin{document}
\setlength\baselineskip{20pt}

\preprint{\tighten \vbox{\hbox{CALT-68-2210}
        \hbox{nucl-th/9901064} 
}}

\title{ Radiation pions in two-nucleon effective field theory  \\  }

\author{Thomas Mehen\footnote{mehen@theory.caltech.edu} and
Iain W.\ Stewart\footnote{iain@theory.caltech.edu} \\[4pt]}
\address{\tighten California Institute of Technology, Pasadena, CA, USA 91125 }

\maketitle

{\tighten
\begin{abstract}

For interactions involving two or more nucleons it is useful to divide pions into
three classes: potential, radiation, and soft.  The momentum threshold for the
production of radiation pions is $Q_r = \sqrt{M_N m_\pi}$.  We show that radiation
pions can be included systematically with a power counting in $Q_r$. The leading
order radiation pion graphs which contribute to NN scattering are evaluated in the
PDS and OS renormalization schemes and are found to give a small contribution. 
The power counting for soft pion contributions is also discussed.

\end{abstract}
}
\vspace{0.7in}

\newpage

Effective field theory is a useful method for describing two-nucleon systems
\cite{EFT}.  Recently, Kaplan, Savage, and Wise (KSW) \cite{ksw1,ksw2} introduced
a power counting which accounts for the effect of large scattering lengths. A
similar power counting is discussed in Ref.~\cite{Bira}. According to the KSW
power counting, the leading order calculation involves only a dimension 6,
four-nucleon operator which is treated nonperturbatively. Higher derivative
operators and pion exchange are treated perturbatively.  In evaluating diagrams
with pions, three types of contributions can be identified: potential, radiation, and
soft.  Since these pion effects differ in size, they each have a different power
counting.  This distinction arises because there are several scales associated with
two nucleon systems.  In this respect the theory is similar to NRQCD and NRQED
\cite{NRQCD}.  

In NRQCD there are three mass scales associated with non-relativistic systems
containing two heavy quarks: the heavy quark mass $M$, momenta $\sim M v$, and
kinetic energy $\sim M v^2$, where $v$ is the relative velocity.  QCD effects at the
scale $M$ are integrated out and appear as local operators in NRQCD.  The
remaining low energy contributions can be divided into potential, radiation
(sometimes referred to as ultra-soft), and soft pieces
\cite{lm,gr,ls,labelle,beneke,gries}.  Potential gluons have energy of order $M v^2$
and momentum of order $M v$, radiation gluons have energy and momentum of
order $M v^2$, and soft gluons have energy and momentum of order $M v$.  The
power counting for radiation gluons requires the use of a multipole expansion at a
quark-gluon vertex \cite{gr,labelle}.  The $v$ power counting of potential and
radiation gluons can be implemented in the effective Lagrangian by introducing
separate gluon fields and rescaling the coordinates and fields by powers of $v$
\cite{lm,ls}.  In Ref.~\cite{beneke} the separation of scales was achieved on a
diagram by diagram basis using a threshold expansion.  The potential, radiation,
and soft regimes were shown to correctly reproduce the low energy behavior of
relativistic diagrams in a scalar field theory.  In Ref.~\cite{gries} it was pointed out
that these effects may be reproduced by an effective Lagrangian in which separate
fields are also introduced for the soft regime.  Note that soft contributions come
from a larger energy scale than potential and radiation effects.  The heavy quark
system does not have enough energy to radiate a soft gluon, so they only appear
in loops.  In Ref.~\cite{beneke} it was shown that soft contributions to scattering
do not appear until graphs with two or more gluons are considered.  

In the nucleon theory there is another scale because the pions are massive.  For
the purpose of power counting it is still useful to classify pion contributions as
potential, radiation, or soft.  For a pion with energy $q_0$ and momentum $q$, a
potential pion has $q_0\sim q^2/M$ where $M$ is the nucleon mass, while a
radiation or soft pion has $q_0\sim q \ge m_\pi$.  In non-relativistic theory,
integrals over loop energy are performed via contour integration.  Potential pions
come from contributions from the residue of a nucleon pole and give the
dominant contribution to pion exchange between two nucleons.  For these pions,
the energy dependent part of the pion propagator is treated as a perturbation
because the loop energy, $q_0\sim q^2/M \ll q$.  The residues of pion propagator
poles give radiation or soft pion contributions.  The power counting for soft and
radiation pions differs.  For instance, the coupling of radiation
pions to nucleons involves a spatial multipole expansion, while the coupling to
soft pions does not.  

In section A, a power counting for graphs with radiation pions is given. To illustrate
this power counting, we compute the leading radiation pion contribution to S-wave
nucleon-nucleon scattering.  The power counting for soft pion contributions is
discussed in section B, and an illustrative example is given.  We emphasize the
importance of not double counting when adding radiation and soft pion 
contributions.

\subsection{Radiation Pions}

In chiral perturbation theory the expansion is in powers of momenta and the pion
mass $m_\pi$.  For power counting potential pions it is convenient to take the
nucleon momentum $p=M v\sim m_\pi$ \cite{ksw2,lm}, so $v=m_\pi/M\sim 0.15$. 
The situation is different for radiation pions.  There is a new scale associated with
the threshold for pion production, which occurs at energy $E=m_\pi$ in the center
of mass frame. This corresponds to a nucleon momentum $p=Q_r$, where
$Q_r\equiv \sqrt{M m_\pi}=360\,{\rm MeV}$.  Because the radiation pion fields
cannot appear as on-shell degrees of freedom below the threshold $E= m_\pi$,
one expects that the radiation pion can be integrated out for $p \ll Q_r$. (Potential
pions should be included for $p\gtrsim m_\pi/2$.)  Another way to see that
radiation pions require $p\sim Q_r$ is to note that in order to simultaneously
satisfy $k_0^2= k^2 + m_\pi^2$ and $k_0\sim k\sim M v^2$ requires $v\sim
\sqrt{m_\pi/M} \sim 0.38$.  

The full theory with pions has operators in the Lagrangian with powers of $m_q$
which give all the $m_\pi$ dependence.  If the radiation pions are integrated out,
then the chiral expansion is no longer manifest because there will be $m_\pi$
dependence hidden in the coefficients of operators in the Lagrangian.  One is still
justified in considering the same Lagrangian, but predictive power is lost since it
is no longer clear that chiral symmetry relates operators with a different number
of pion fields.  Also, the $m_\pi$ dependence induced by the radiation pions may
affect the power counting of operators. For example, as shown in Ref.\cite{bkm},
integrating out the pion in the one-nucleon sector induces a nucleon electric
polarizability $\alpha_E \propto 1/m_\pi$.  Alternatively, one can keep chiral
symmetry manifest by working with coefficients in the full theory and including
radiation pion graphs.  This is the approach we will adopt.

The presence of the scale $Q_r$ modifies the power counting of the theory with
radiation pions.  In the KSW power counting, one begins by taking external
momenta $p \sim m_\pi \sim Q$. The theory is organized as an expansion in
powers of $Q$.  To estimate the size of a graph, loop 3-momenta are taken to be
of order $Q$.  However, potential loops within graphs with radiation pions can
actually be dominated by three momenta of order $Q_r$.  To see how this comes
about, consider as an example the graph shown in Fig. 1c.  Let $q$ be the
momentum running through the pion propagator, and let $k$ be the loop
momentum running through a nucleon bubble inside the radiation pion loop. 
The poles from the pion propagator are
\begin{eqnarray}
 {i\over q_0^2-{\vec q}\,^2 -m_\pi^2+i\epsilon} = {i\over
 (q_0-\sqrt{{\vec q}\,^2+m_\pi^2}+i\epsilon)(q_0+\sqrt{{\vec
 q}\,^2+m_\pi^2}-i\epsilon)} \,,
\end{eqnarray}
so the radiation pion has $|q_0| \ge m_\pi$. This energy also goes into the
nucleon bubbles. The $k$ integrand is largest when the nucleons are close to
their mass shell. But since the energy going into the loop is $\sim m_\pi$, this
occurs when $k^2/M \sim m_\pi$, i.e., $k \sim Q_r$.  We will begin by considering
the contribution of radiation pions to elastic nucleon scattering at the threshold,
$E = m_\pi$.  At this energy, external and potential loop momenta are of the
same size and power counting is easiest. Because $p \sim Q_r$ it is obvious that
we want to count powers of $Q_r$ rather than $Q$.
 
Before discussing the power counting, recall the Lagrangian with pions 
and nucleons \cite{ksw2}:
\begin{eqnarray}\label{Lag}
{\cal L}_\pi &=& \frac{f^2}{8} {\rm Tr}\,( \partial^\mu\Sigma\: \partial_\mu
\Sigma^\dagger )+\frac{f^2\omega}{4}\, {\rm Tr} (m_q \Sigma+m_q 
\Sigma^\dagger)+ \frac{ig_A}2\, N^\dagger \sigma_i (\xi\partial_i\xi^\dagger -
\xi^\dagger\partial_i\xi) N  \nn\\ &+& N^\dagger \bigg( i
D_0+\frac{\vec D^2}{2M} \bigg) N - {C_0^{(s)}} ( N^T P^{(s)}_i
N)^\dagger ( N^T P^{(s)}_i N) \label{Lpi} \\ &+&   {C_2^{(s)}\over
8} \left[ ( N^T P^{(s)}_i N)^\dagger ( N^T P^{(s)}_i
\:\tensor{\nabla}^{\,2} N) + h.c. \right] -{D_2^{(s)}} \omega {\rm
Tr}(m^\xi ) ( N^T P^{(s)}_i N)^\dagger ( N^T P^{(s)}_i N) + \ldots
\nn .
\end{eqnarray}
Here $g_A=1.25$ is the nucleon axial-vector coupling, $\Sigma = \xi^2$ is the
exponential of pion fields, $f=131\, {\rm MeV}$ is the pion decay constant,
$m^\xi=\frac12(\xi m_q \xi + \xi^\dagger m_q \xi^\dagger)$, where $m_q={\rm
diag}(m_u,m_d)$ is the quark mass matrix, and $m_\pi^2 = w(m_u+m_d)$.  The
matrices $P_i^{(s)}$ project onto states of definite spin and isospin, and the
superscript $s$ denotes the partial wave amplitude mediated by the operator. 
This paper will be concerned only with S-wave scattering, so $s={}^1\!S_0,
{}^3\!S_1$.

The $C_0$ operator mediates S-wave nucleon transitions.  The $D_2$ operator is
important because it will be necessary to introduce counterterms proportional to
$m_\pi^2$ to regulate UV divergences appearing in the graphs evaluated below. 
The parameters appearing in Eq.~(\ref{Lag}) are bare parameters which require
renormalization. For systems with two or more nucleons, it is necessary to
introduce finite subtractions in order to obtain manifest power counting. Two such
renormalization schemes are Power Divergence Subtraction (PDS) \cite{ksw1,ksw2}
and a momentum subtraction scheme (OS) \cite{ms0}.  These renormalization
schemes are discussed extensively in Ref.\cite{ms1}.  The renormalized coefficient
$C_0(\mu_R)\sim 4\pi/(M\mu_R)$ where $\mu_R$ is the renormalization point.  A
loop with two nucleon propagators gives $\sim M p/(4\pi)$, so for $\mu_R\sim p$
the $C_0(\mu_R)$ bubble graphs should be summed.  For a scattering length $a$,
this summation includes powers of $(a\,p)$ to all orders\cite{ksw1}, as desired
since $a$ is large. The coefficients $C_2(\mu_R)$ and $D_2(\mu_R)$ scale as $\sim
1/\mu_R^2$, so for $\mu_R\sim p \sim m_\pi$ they are treated perturbatively.  The
ellipsis in Eq.~(\ref{Lpi}) denotes higher order terms including contact interactions
which for NN scattering begin to contribute at next-to-next-to-leading order
(NNLO).  These will not be considered here.

Next, we introduce the power counting at the scale $Q_r$. A scheme with manifest
power counting will be used, so that $C_0(\mu_R) \sim 1/(M \mu_R), C_2(\mu_R)
\sim 1/(M\Lambda \mu_R^2)$, etc., where $\Lambda$ is the range of the theory. We
will take $p \sim \mu_R \sim Q_r$.  A radiation loop has $q_0\sim q \sim m_\pi$ so
$d\,^4q\sim Q_r^8/M^4$, where $q$ is the momentum running through the pion
propagator.  A radiation pion propagator gives a $M^2/Q_r^4$, while the derivative
associated with a pion-nucleon vertex gives $Q_r^2/M$. A nucleon propagator
gives a $M/Q_r^2$.  External energies and momenta are kept in the nucleon
propagator since $E\sim p^2/M \sim Q_r^2/M$.  Furthermore, it is appropriate to
use a multipole expansion for radiation pion-nucleon vertices which is similar to
the treatment of radiation gluons in NRQCD \cite{gr}.  Therefore, radiation pions
will not transfer three-momenta to a nucleon. This is usually equivalent to
expanding in powers of a loop momentum divided by $M$ before doing the loop
integral. A potential loop will typically have running through it either an external or
radiation loop energy $\sim Q_r^2/M$.  Therefore, in these loops the loop energy
$k_0\sim Q_r^2/M$, while the loop three momentum $k\sim Q_r$, so $d^4k\sim
Q_r^5/M$.  It is not inconsistent for $k\sim Q_r$ while $q\sim Q_r^2/M$, since three
momenta are not conserved at the nucleon-radiation pion vertices.   At the scale
$Q_r$ potential pion propagators may still be treated in the same way,
$i/(k_0^2-k^2-m_\pi^2)= -i/(k^2+m_\pi^2)+ {\cal O}(k_0^2/k^4)$, which has an
expansion in $Q_r^2/M^2$.  We will see through explicit examples that this power
counting correctly estimates the size of radiation pion graphs.

Note that only the potential loop measure gives an odd power of $Q_r$, so
without potential loops the power counting reduces to power counting in powers
of $m_\pi$. The power counting here therefore correctly reproduces the usual
chiral power counting used in the one nucleon sector\cite{jm}. 

\begin{figure}[!t]
  \centerline{\epsfysize=7.5truecm \epsfbox{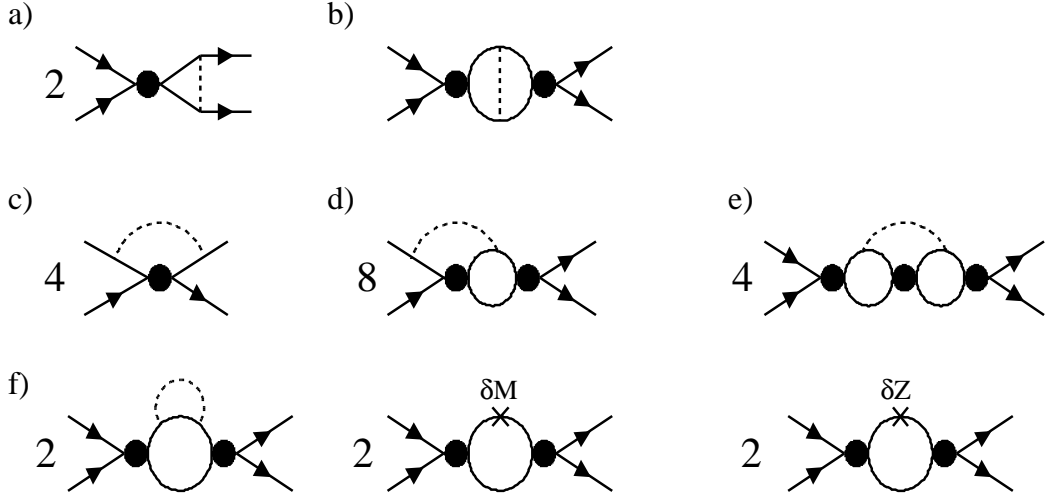}  }
 {\tighten
\caption[1]{Leading order radiation pion graphs for $NN$ scattering.  The solid
lines are nucleons, the dashed lines are pions and $\delta M$, $\delta Z$ are the
mass and field renormalization counterterms. The filled dot denotes the
$C_0(\mu_R)$ bubble chain.  There is a further field renormalization contribution
that is calculated in text, but not shown.} \label{Qr3} }
\end{figure}

Graphs with one radiation pion and additional higher order contact interactions
or potential pions are suppressed by factors of $Q_r/\Lambda$ relative to graphs
with a single radiation pion and $C_0$ vertices. The $Q_r$ expansion is a chiral
expansion about $m_\pi=0$, so there is a limit of QCD where it is justified.  The
scale $\Lambda$ is unknown. One possible estimate is $\Lambda_{\rm NN}=8\pi
f^2/(M g_A^2)= 300\,{\rm MeV}$ since a graph with $m+1$ potential pions is
suppressed by $p/300\,{\rm MeV}$ relative to a graph with $m$ potential pions.
However, this order of magnitude estimate only takes into consideration a partial
subset of the graphs of the theory. As argued in Ref.~\cite{ms1}, it is possible
that the range is of order the scale of short range interactions that are integrated
out, implying $\Lambda \sim 500\,{\rm MeV}$. In fact, the accuracy of NLO
computations of nucleon-nucleon phase shifts is in agreement with this
physically motivated estimate of the range.  We will assume that an expansion in
$Q_r/\Lambda$ is valid.  This hypothesis will be tested further by seeing how
well the effective theory makes predictions at $p\sim 300\,{\rm MeV}$.  For
example, processes with external pions could be considered.   If the $Q_r/
\Lambda$ expansion is not convergent, then application of the theory is
restricted to $p < Q_r$. 

The radiation pion graphs that give the leading order contribution to
nucleon-nucleon scattering are shown in Fig.~\ref{Qr3}. The filled dot denotes
the leading order interaction between nucleons, a $C_0(\mu_R)$ bubble sum. 
We illustrate the power counting with an example, the graph in Fig. 1d. For the
moment, replace the $C_0$ bubble sums with single $C_0$ vertices.  Each $C_0$
gives a factor of $1/M Q_r$ and each nucleon line gives a factor $M/Q_r^2$. The
derivatives from the pion couplings combine with the radiation pion propagator
to give a factor of unity. The radiation loop gives $Q_r^8/M^4$, while the nucleon
bubble loop gives $Q_r^5/M$. There is also a factor of $1/f^2$ from pion
exchange, and two factors of $1/4 \pi$ from the radiation loop giving a
$1/\Lambda_\chi^2$.  ($\Lambda_\chi \sim 1 \,{\rm GeV}$ is the chiral symmetry
breaking scale.) Combining all factors, we find that this graph scales like
$Q_r^3/(M^3 \Lambda_\chi^2)$. This graph is suppressed relative to the leading
order amplitude, $A^{(-1)}$, by a factor of $Q_r^4/(M^2 \Lambda_\chi^2) =
m_\pi^2/\Lambda_\chi^2$.  Note that $C_0$ bubbles are summed on external
nucleon lines as well as in the interior of the radiation loop, and each graph in
the sum has the same size.  It is straightforward to verify that all graphs in Fig.  1
scale the same way. For external bubble sums we can simply use the vertex $i
A^{(-1)}$ where $A^{(-1)}$ is the leading order S-wave amplitude,
\begin{eqnarray}\label{LO}
  A^{(-1)} = -{4\pi \over M} {1 \over \gamma+ip}  \,,
\end{eqnarray}
and the pole $\gamma = 4\pi/MC_0(\mu_R) + \mu_R \sim 1/a$. 
Graphs with two radiation pions are suppressed by at least 
$Q_r^8/(M^4 \Lambda_\chi^4) = m_\pi^4/\Lambda_\chi^4$ and will not be
considered.

The first graphs we consider are those in Fig.~\ref{Qr3}a,b. These graphs
have contributions from potential and radiation pions, and it may not
be obvious that a clean separation occurs. Here the energy integrals will be
evaluated without any approximations, after which the graphs split into
radiation and potential parts.  The graph in Fig.~\ref{Qr3}a gives:
\begin{eqnarray}\label{oneloop} 
iA^{(-1)} {g_A^2 \over 2 f^2} \int {d^D q \over
  (2 \pi)^D} {i \over \frac{E}2+q_0 -{({\vec q}-{\vec p})^2 \over 2M} +i
  \epsilon}\: {i \over \frac{E}2-q_0 -{({\vec q}-{\vec p})^2 \over 2M} +i
  \epsilon}\: {i\: {\vec q\,}^2 \over q_0^2 - {\vec q}\,^2 - m_\pi^2 + i \epsilon} \,.  
\end{eqnarray} 
(Throughout this paper we will include a factor of $(\mu/2)^{4-D}$ in the loop
measures.) Performing the $q_0$ integral gives a term from the residue of the
nucleon pole and a term from the pion pole, 
\begin{eqnarray}
&& -iA^{(-1)} {g_A^2 \over 2 f^2} \int {d^n q \over (2 \pi)^n} 
{M \over {({\vec q}-{\vec p})^2 }-M E}\: {{\vec q}\,^2 \over {\vec q}\,^2 + 
m_\pi^2 -[\frac{E}2-{({\vec q}-{\vec p})^2 \over 2M}]^2 }\: \label{bme}
\\ && -iA^{(-1)} {g_A^2 \over 4 f^2} \int {d^n q \over (2 \pi)^n} {{\vec q}\,^2
\over \sqrt{{\vec q}\,^2 + m_\pi^2}}\: {1 \over {E\over 2}+\sqrt{{\vec q}\,^2 
+ m_\pi^2}-{({\vec q}-{\vec p})^2 \over 2M} }\: 
{1 \over {E\over 2}-\sqrt{{\vec q}\,^2 + m_\pi^2}-{({\vec q}-{\vec p})^2 
\over 2M}} \,, \label{bme2} 
\end{eqnarray} 
where $n=D-1$.  Eq.~(\ref{bme}) is the potential pion contribution. Expanding in
$[\frac{E}2-{({\vec q}-{\vec p})^2 \over 2M}]^2=[{ 2{\vec q}.{\vec p}-{\vec q}\,^2
\over 2M}]^2$ gives the result in Ref.  \cite{ksw1,ksw2}.  The subleading terms in
this expansion are suppressed by\footnote{ \tighten Note that $m_\pi^2/q^2\sim
m_\pi/M$, but we have kept the $m_\pi^2$ term in the potential pion propagator in
Eq.~(\ref{bme}).  We could consider expanding in $m_\pi/q$ using the asymptotic
expansion techniques discussed in section B, but for $p\sim m_\pi$ these terms
would have to be resummed.  Unlike the soft and radiation contributions, there is
no issue of double counting for potential pions, so for simplicity we will simply
keep the $m_\pi^2$ in the propagator.} $m_\pi^2/M^2$.  Eq.~(\ref{bme2}) is the
radiation pion contribution.  With $|\vec q|<M$, we may take $({\vec q}-{\vec
p})^2/M\to p^2/M$ in the last two propagators, which is the same approximation
that is made by performing the multipole expansion.  Finally, we use the equations
of motion to set $E - p^2/M = 0$. It is important to note that we have not neglected
$E$ relative to $|{\vec q}|$.  For $n=3-2\epsilon$, Eq.~(\ref{bme2}) becomes 
\begin{eqnarray}\label{gr1} 
a) =\, i A^{(-1)} {g_A^2 \over 4 f^2} \int {d^n q \over (2 \pi)^n}{{\vec q\,}^2 \over 
  ({\vec q}\,^2 +  m_\pi^2)^{3/2}} = 
  -3 i A^{(-1)} { g_A^2 m_\pi^2\over (4 \pi f)^2} \bigg[ {1 \over \epsilon} + 
  {1\over 3} - {\rm ln}\Big({m_\pi^2 \over {\overline{\mu}}^2}\Big) \bigg] \,, 
\end{eqnarray} 
where $\overline{\mu}^2=\mu^2 \pi e^{-\gamma_E}$.  Note that this integral is 
finite in three dimensions ($n=2$).

The next graph we consider is shown in Fig.~\ref{Qr3}b.  We have chosen to
route loop momenta so that $q$ runs through the pion and $\pm k$ and
$\pm(k+q)$ run through the nucleon lines. The momentum $k$ is potential, while
$q$ can be potential or radiation. Doing the $k_0$ contour integral and
combining the two terms gives:
\begin{eqnarray}
&& -2\, [A^{(-1)}]^2\, {g_A^2 \over 2 f^2} \int {d^{n} k \over (2
 \pi)^{n}} \int {d^D q \over (2 \pi)^D} {{\vec q}\,^2\ \over q_0^2 -
 {\vec q}\,^2 - m_\pi^2 + i \epsilon}\: {1 \over E - {{\vec k}^2
 \over M}+ i\epsilon}\: {1 \over E - {({\vec k} + {\vec q})^2 \over
 M}+i \epsilon} \\ &&\qquad\qquad\qquad\qquad\qquad \times {
 E-{({\vec k}+{\vec q})^2 \over 2M}-{{\vec k}\,^2\over 2M} \over
 [E-{({\vec k}+{\vec q})^2 \over 2M}-{{\vec k}\,^2\over 2M}-q_0+ i
 \epsilon] [E-{({\vec k}+{\vec q})^2 \over 2M}-{{\vec k}\,^2\over
 2M}+q_0+ i \epsilon]} \nn \,.
\end{eqnarray}
Doing the $q_0$ integral gives two terms, but the radiation and
potential contributions are still mixed.  Combining these gives
\begin{eqnarray}
&& i\, [A^{(-1)}]^2\, {g_A^2 \over 2 f^2} \int {d^{n} k \: d^nq
\over (2 \pi)^{2n}} {{\vec q}\,^2\ \over \sqrt{{\vec q}\,^2
 + m_\pi^2} }\: {1 \over E - {{\vec k}^2 \over M}}\: {1 \over E -
 {({\vec k} + {\vec q})^2 \over M}}\: { 1 \over E-{({\vec
k}+{\vec q})^2 \over 2M}-{{\vec k}\,^2\over 2M}-\sqrt{{\vec q}\,^2
 + m_\pi^2} }   \,,
\end{eqnarray}
which can be split into potential and radiation parts
\begin{eqnarray}\label{twoloop}
i [A^{(-1)}]^2\, {g_A^2 \over 2f^2} \int {d^n k \over (2 \pi)^n}
\int {d^n q \over (2 \pi)^n} {{\vec q}\,^2 \over \sqrt{ {\vec
q}\,^2 + m_\pi^2 }}\: {1 \over E - {{\vec k}^2 \over M}}\: {1
\over \sqrt{ {\vec q}\,^2 + m_\pi^2}- {(2{\vec k}\cdot{\vec q} +
{\vec q}\,^2)\over 2 M}  }
\\ \times\Bigg[ {-1 \over E - {({\vec k}+{\vec q})^2 \over M}}
+{1 \over E - {{\vec k}^2 \over M} - {(2{\vec k}\cdot{\vec q} +
 {\vec q}\,^2)\over 2 M}-\sqrt{{\vec q}\,^2 + m_\pi^2}} \Bigg] \nn \,.
\end{eqnarray}
The first term in Eq.~(\ref{twoloop}) is the two-loop potential pion graph
evaluated in Ref.\cite{ksw2}. The factors of $(2{\vec k}\cdot{\vec q} + {\vec
q}\,^2)/(2M)$ appearing in the denominators can be dropped because the loop
integral is dominated by $k,q \ll M$ and therefore $(2{\vec k} \cdot {\vec q} + {\vec
q}\,^2)/(2M) \ll \sqrt{{\vec q}\,^2 + m_\pi^2}$.  For the second term, which is the
radiation pion contribution, this is equivalent to the multipole expansion. 
Momenta $k\sim \sqrt{M m_\pi}$ and $q\sim m_\pi$  dominate the integrals in the
second term.  In Ref.~\cite{beneke}, the potential and radiation parts of the graph
in Fig.~\ref{Qr3}b were evaluated in the limit $m_\pi=0$, and shown to correctly
make up the corresponding part of the fully relativistic calculation.  The
calculation in Ref.~\cite{beneke} agrees with Eq.~(\ref{twoloop}) for $m_\pi=0$. 
Note that the radiation part would not agree if we assumed $k\sim m_\pi$ and
used static nucleon propagators in the radiation loop\footnote{\tighten
Furthermore, if static nucleon propagators are used one obtains a linear 
divergence requiring a non-analytic counterterm $\propto m_\pi$ \cite{gg}.}.  
For $n=3-2\epsilon$ the radiation part of Eq.~(\ref{twoloop}) is
\begin{eqnarray}\label{gr2}
b) &=&\, i [A^{(-1)}]^2\, {g_A^2 \over 2f^2} \int {d^n k \over (2
\pi)^n} \int {d^n q \over (2 \pi)^n} {{\vec q}\,^2 \over {\vec
q}\,^2 + m_\pi^2 }\: {1 \over E - {{\vec k}\,^2 \over M}}\: {1
\over E - {{\vec k}\,^2 \over M} -\sqrt{{\vec q}\,^2 + m_\pi^2}}
\\ &=&[A^{(-1)}]^2\, {g_A^2 M m_\pi^2 \over (4 \pi f)^2}\, \Bigg\{
{3\,p \over 4 \pi}\, \bigg[ {1 \over \epsilon} + {7\over 3} -
2\,{\rm ln}\, 2 - {\rm ln}\Big({m_\pi^2 \over {\overline{\mu}}^2}
\Big) - {\rm ln}\Big({-p^2 \over {\overline{\mu}}^2}\Big) \bigg] +
{i\sqrt{M m_\pi} \over 4\sqrt{\pi}}\,I_1 \Big({E\over m_\pi}\Big)
\Bigg\} \,, \nn
\end{eqnarray}
where
\begin{eqnarray}\label{I1}
I_1(x) = {3 \over 2} {\Gamma(-{5\over 4}) \over \Gamma({5 \over
  4})} \, {}_3 F_2 \bigg( \{-{5\over 4},-{1\over 4},{1\over 4}\}, \{
  {1\over 2},{5\over 4}\},x^2 \bigg) + {x \Gamma({1 \over 4})\over
  \Gamma({7 \over 4})}  \, {}_3 F_2 \bigg( \{-{3\over 4},{1\over
  4},{3\over 4}\}, \{ {3\over 2},{7\over 4}\} ,x^2 \bigg)\,.
\end{eqnarray}
For $n=2$ the loop integral in Eq.~(\ref{gr2}) is finite, except for the function $I_1$
which has a $p^2/(n-2)$ pole.  In PDS this pole would effect the running of
$C_2(\mu_R)$, but we will see that contributions proportional to $I_1$ will cancel
in the sum of graphs in Fig.~\ref{Qr3}. Since $A^{(-1)}\sim 1/(M Q_r)$ the results in
Eqs.~(\ref{gr1}, \ref{gr2}) are order $Q_r^3/(M^3 \Lambda_\chi^2)$ as expected. At
one-loop the $1/\epsilon$ pole in Eq.~(\ref{gr1}) is cancelled by a
counterterm\footnote{\tighten The bare coefficients are written as $C^{\rm
bare}=\delta^{\rm uv}C + C^{\rm finite}$. In OS and PDS additional finite
subtractions are made so that $C^{\rm finite}= C(\mu_R)-\sum \delta^n C(\mu_R)$,
see Ref.~\cite{ms1}.}
\begin{eqnarray}\label{cntr}
   \delta^{{\rm uv},1a}D_2=-3\, C_0^{\rm finite}\,{g_A^2 \over (4
   \pi f)^2} \bigg[ {1 \over \epsilon}-\gamma_E +\ln{(\pi)} \bigg]  \,.
\end{eqnarray}
For higher loops the $1/\epsilon$ poles in Eq.~(\ref{gr1}), Eq.~(\ref{gr2}), and
$\delta^{{\rm uv},1a}D_2$ dressed with $C_0$ bubbles cancel. Note that the
$O(\epsilon)$ piece of the bubbles give a finite contribution,
\begin{eqnarray}
  -C_0(\mu_R) \left( {M p \over 4 \pi} \right) \left\{ 1 +\epsilon
  \left[ 2-2\,{\rm ln}\,2 - {\rm ln} \left( {-p^2-i\epsilon \over
  {\overline{\mu}}^2}\right) \right] \right\} \,.
\end{eqnarray}
The result of combining Figs.~\ref{Qr3}a,b and $\delta^{{\rm uv},1a}D_2$ dressed
by $C_0$ bubbles is:
\begin{eqnarray}\label{pt1}
a)+b)=\,  3i\,[A^{(-1)}]^2\,{ g_A^2 m_\pi^2 \over (4 \pi f)^2}
  \,{M\gamma\over 4\pi}\,\bigg[ {1\over 3} - {\rm ln}\Big({m_\pi^2
   \over \mu^2}\Big)\bigg]  + i\,[A^{(-1)}]^2\,{ g_A^2 M m_\pi^2
   \over (4 \pi f)^2} {\sqrt{M m_\pi} \over 4\sqrt{\pi}}\, I_1
   \Big({E\over m_\pi}\Big) \,.
\end{eqnarray}

Next we consider the graphs in Fig.~\ref{Qr3}c,d,e. The loop integrals in these
graphs vanish if the pion pole is not taken so there is no potential pion
contribution. As pointed out in Ref.~\cite{ksw2}, emission of the radiation pion in
these graphs changes the spin/isospin of the nucleon pair. Therefore, if the
external nucleons are in a spin-triplet(singlet) state, then the coefficients
appearing in the internal bubble sum are $C_0^{(^1S_0)}(\mu_R)\:
(C_0^{(^3S_1)}(\mu_R))$.  The notation $C_0$ ($C_0^{\,\prime}$) will be used for
vertices outside (inside) the radiation pion loop. We begin with Fig.~\ref{Qr3}c. 
The contribution from the graph with $m$ nucleon bubbles in the internal bubble
sum is
\begin{eqnarray}\label{gr3a}
 && 4\,{g_A^2 \over 2 f^2} \int {d^D q \over (2
  \pi)^D} {i \over q_0 - i \epsilon}\: {i \over q_0 - i \epsilon}\:
  {-i \,{\vec q}\,^2 \over q_0^2 -{\vec q}\,^2 - m_\pi^2 +
  i\epsilon} [- i C_0^{\,\prime}(\mu_R)]^{m+1} \nonumber \\ 
 &&\qquad\qquad \times
  \left[ \int {d^D k \over (2 \pi)^D} {i \over -k_0 -q_0
  +\frac{E}2-{({\vec k}-{\vec q})^2 \over 2M} +i \epsilon}\: {i
  \over k_0 +\frac{E}2-{{\vec k}^2 \over 2M} +i \epsilon} \right]^m \,,
\end{eqnarray}
where we used the multipole expansion and then the equations of motion to
eliminate $E$ and $p$ from the first two propagators.  All nucleon propagators
have a $q_0$ pole above the real axis, while the pion propagator has one pole
above and one below.  Therefore, the $q_0$ contour is closed below. The $d^D
k$ integrals are also easily performed giving
\begin{eqnarray} \label{gphcm}
&& {-i}{ g_A^2 C_0^{\,\prime}(\mu_R) \over f^2}\left[ {-C_0^{\,\prime}(\mu_R) M
\Gamma({1-n/2}) \over (4 \pi)^{n/2}}\right]^m\! \int\! {d^n q \over (2
\pi)^n} {{\vec q}\,^2 \left[(-p^2 +M \sqrt{{\vec q}\,^2 +
m_\pi^2}\, )^{n/2-1}-\mu_R\right]^m\over ({\vec q}\,^2 +
m_\pi^2)^{3/2} } \,. \nn \\
&& 
\end{eqnarray}
Note  that the size of the loop momenta $k$ in the nucleon bubbles is $\sim \sqrt{M
m_\pi}$ even for $p < \sqrt{M m_\pi}$. The $\mu_R$ inside the brackets comes from
inclusion of the PDS or OS $\delta^n C_0(\mu_R)$ counterterm graphs for the
internal bubble sum. The integral will be dominated by ${\vec q} \sim m_\pi$ so the
graph will scale as
\begin{eqnarray}
{1\over \Lambda_\chi^2}\:{m_\pi^2 \over M
\mu_R}\bigg({\sqrt{M m_\pi} \over \mu_R }\bigg)^m \,.
\end{eqnarray}
Since $\mu_R\sim \sqrt{M m_\pi}$, all graphs in the sum are of order
$Q_r^3/(M^3 \Lambda_\chi^2)$.

For Figs.~\ref{Qr3}c,d,e the sum over bubbles should be done before the radiation
loop integral. The reason is that an arbitrary term in the bubble sum has a much
different dependence on the energy flowing through it than the sum itself. This can
be seen in the $\vec q$ dependence in Eq.~(\ref{gphcm}). If we integrate over $\vec
q$ then terms in the sum may diverge whereas the integral of the complete sum is
finite.  In fact, for $n=3$, Eq.~(\ref{gphcm}) has divergences of the form
$\Gamma(-1-m/4) {\rm F}(E^2/m_\pi^2)$ and $\Gamma(-1/2-m/4) E\, {\rm
F}(E^2/m_\pi^2)$ where ${\rm F}$ is a hypergeometric function. These divergences
are misleading because for momenta $>1/a$ we know that the correct form of the
leading order four point function falls off as $1/p$.  For this reason the summation
is performed before introducing counterterms to subtract divergences. (This
approach is also taken in the analysis of three body interactions in
Ref.~\cite{3bdy}). Summing over $m$, Eq.~(\ref{gphcm}) becomes:
\begin{eqnarray}\label{gr3}
c) &=&\, {- i}\, {g_A^2 \over f^2} {4 \pi \over M} \int {d^{n} q \over (2
   \pi)^{n}}{{\vec q}\,^2 \over ({\vec q}\,^2 + m_\pi^2)^{3/2}}\: { 1
   \over \gamma' - \left[-p^2 +M \sqrt{{\vec q}\,^2 +
   m_\pi^2}\right]^{n/2-1}}  \nn\\
&=&\,  {ig_A^2 \over \sqrt{\pi} f^2}  \Big({m_\pi \over M}\Big)^{3/2} 
    I_2 \Big( {E \over m_\pi} \Big)\,,
\end{eqnarray}
where $\gamma'=4\pi/MC_0^{\,\prime}(\mu_R)+\mu_R \sim 1/a$. As expected the
graph scales as $Q_r^{3}$. In the limit $n\to3$, $I_2$ is finite and given by
\begin{eqnarray}\label{I2}
I_2(x) &=&  {\Gamma(-{3\over 4}) \over \Gamma({3 \over 4})} \,
{}_3 F_2 \left( \{-{3\over 4},{1\over 4},{3\over 4}\}, \{ {1\over
2},{7\over 4}\},x^2 \right) - {3 x \over 2} {\Gamma({3 \over
4})\over \Gamma({9 \over 4})} \, {}_3 F_2 \left( \{-{1\over
4},{3\over 4},{5\over 4}\}, \{ {3\over 2},{9\over 4}\} ,x^2 \right
)  \nn \\ &&\quad + {\cal O}(\gamma'/\sqrt{Mm_\pi}) \,.
\end{eqnarray}
$I_2$ is manifestly $\mu_R$ independent and is also finite as
$n\to 2$. 

Next we consider the graph in Fig.~\ref{Qr3}d. Integrals are done
in the same manner as that of Fig.~\ref{Qr3}c. For
$n=3-2\epsilon$, Fig.~\ref{Qr3}d is
\begin{eqnarray}\label{gr4}
d) &=& {-4i\, A^{(-1)} } {g_A^2 \over 2 f^2}{\Gamma(n/2-1)\over  (4\pi)^{n/2-1}} 
 \int{d^{n} q \over (2 \pi)^{n}}{{\vec q}\,^2 \over ({\vec q}\,^2 + m_\pi^2)^{3/2}} {
  \left(-p^2 +M \sqrt{{\vec q}^2 + m_\pi^2}\,\right)^{n/2-1}-(-p^2)^{n/2-1} \over
  {\gamma'} - \left(-p^2 +M \sqrt{{\vec q}\,^2 + m_\pi^2}\,\right)^{n/2-1}}\nn \\
&=&-12 i\, A^{(-1)} { g_A^2 m_\pi^2 \over (4 \pi f)^2} \bigg[{1\over
  \epsilon} + {1\over 3} - {\rm ln}\Big({m_\pi^2 \over {\overline \mu}^2}\Big) \bigg]
  -4 {(p-i\gamma')\over \sqrt{\pi}} {M A^{(-1)} \over 4 \pi} {g_A^2 \over 2 f^2}
  \Big({m_\pi \over M} \Big)^{3/2} I_2 \Big({E\over m_\pi} \Big)\,.
\end{eqnarray}
Fig.~\ref{Qr3}d is finite for $n=2$, except for the function $I_1$. The $1/\epsilon$
pole in Eq.~(\ref{gr4}) is cancelled by a new tree level counterterm $i \delta^{\rm
uv,1d} D_2 m_\pi^2$ where $\delta^{\rm uv,1d}D_2$ has the same form as
Eq.~(\ref{cntr}) except with a $-12$ instead of a $-3$.

Evaluation of Fig.~\ref{Qr3}e is also similar to Fig.~\ref{Qr3}c.
For $n=3-2\epsilon$ we find:
\begin{eqnarray} \label{gr5}
e)&=&{-2\,i }\,{(p-i\gamma')^2\over\sqrt{\pi}}  \bigg[{ M A^{(-1)} \over 4
  \pi}\bigg]^2 {g_A^2 \over 2 f^2} \bigg({m_\pi \over M}\bigg)^{3/2}
  I_2\Big({E \over m_\pi} \Big) + i[A^{(-1)}]^2\, {g_A^2M m_\pi^2
  \over (4 \pi f)^2}\, {\sqrt{M m_\pi} \over 2\sqrt{\pi}}\,
  I_1\left({E \over m_\pi} \right) \nn \\
&&\quad+12\,[A^{(-1)}]^2\, {M p \over 4 \pi}\,{g_A^2 m_\pi^2
  \over (4\pi f)^2}\bigg[{1\over\epsilon}+ {7 \over 3}-2\ln{2} -
  {\rm ln}\Big({m_\pi^2 \over {\overline{\mu}}^2}\Big) - {\rm
  ln}\Big({-p^2 \over {\overline{\mu}}^2}\Big)\bigg] \nn \\
&&\quad -6\, i\,[A^{(-1)}]^2\, {M \gamma' \over 4 \pi}\,{g_A^2 m_\pi^2
  \over (4\pi f)^2}\bigg[{1\over\epsilon}+ {1 \over 3}-
  {\rm ln}\Big({m_\pi^2 \over {\overline{\mu}}^2}\Big) \bigg] \,. 
\end{eqnarray}
This graph is finite for $n=2$, except for the function $I_1$.  A $D_2$ counterterm 
cancels the divergence in the last line,
\begin{eqnarray}  \label{ctgr5}
  \delta^{{\rm uv},1e}D_2=6 \, (C_0^{\rm finite})^2\,{ M\gamma'\over 4\pi} 
  \,{g_A^2 \over (4 \pi f)^2} \bigg[ {1 \over \epsilon}-\gamma_E +\ln{(\pi)} \bigg] \,.
\end{eqnarray}
For two and higher loops the remaining
$1/\epsilon$ poles cancel between Eqs.~(\ref{gr4},\ref{gr5},\ref{ctgr5}) and
$\delta^{{\rm uv},1d}D_2$ dressed with $C_0$ bubbles, so no
new counterterms need to be introduced. The $O(\epsilon)$ piece of the bubbles
again give a finite contribution.  Combining Figs.~\ref{Qr3}c,d,e, and 
$\delta^{{\rm uv},1d}D_2$ and $\delta^{{\rm uv},1e}D_2$ dressed with $C_0$ 
bubbles gives
\begin{eqnarray}\label{ans3}
c)+d)+e) &=&2i\, [ A^{(-1)}]^2 {g_A^2 \over (4 \pi f)^2} \Bigg\{ 6 m_\pi^2\,
{M(\gamma - \gamma'/2)\over 4\pi\,} \bigg[ {1 \over 3} - {\rm ln}\Big({m_\pi^2
\over {\mu}^2}\Big) \bigg]+ { M^{3/2}m_\pi^{5/2} \over 4
\sqrt{\pi}} I_1\Big( {E \over m_\pi} \Big) \nn \\
&&\qquad\qquad\qquad\qquad +{(\gamma - \gamma')^2\over 2\sqrt{\pi}}\,
{(Mm_\pi)^{3/2} \over M}\, I_2\Big({E \over  m_\pi} \Big) \Bigg\}  \,.
\end{eqnarray}

Fig.~\ref{Qr3}f shows a two loop graph with a nucleon self energy on an internal
line. It is important to also include graphs with the one-loop wavefunction and
mass renormalization counterterms, $\delta Z,\delta M$ inserted on the internal
nucleon line. We will use an on-shell renormalization scheme for defining these
counterterms, which ensures that the mass, $M$, appearing in all expressions is
the physical nucleon mass.  The counterterms are:
\begin{eqnarray}
\delta M &=& {3 g_A^2 m_\pi^3 \over 16 \pi f^2}\ , \qquad\qquad \delta Z =
{9\over 2}{g_A^2 m_\pi^2 \over (4 \pi f)^2} \left({1 \over
\epsilon} +{1\over 3} - {\rm ln}\left({m_\pi^2 \over
{\overline{\mu}}^2}\right) \right) \,.
\end{eqnarray}
The result from the graphs in Fig.\ref{Qr3}f is then
\begin{eqnarray}\label{gph6}
 f) = -3i [A^{(-1)}]^2 { g_A^2 \over (4 \pi f)^2} {M^{3/2}m_\pi^{5/2}
\over 4\sqrt{\pi}} I_1 \left({E \over m_\pi} \right) \,.
\end{eqnarray}
When Eq.~(\ref{gph6}) is added to Eqs.~(\ref{pt1},\ref{ans3}) the terms
proportional to $I_1$ cancel. To implement PDS we must consider the value of
the graphs in Fig.~\ref{Qr3}f using Minimal Subtraction with $n=2$. For
$n=2+\epsilon$ we have $\delta M= 3g_A^2 m_\pi^2\mu/(16\pi f^2\,\epsilon)$ and
$\delta Z=0$, which makes the sum of graphs in Fig.~\ref{Qr3}f finite.  Finally,
renormalization of the bare nucleon fields in the Lagrangian, $N^{\rm
bare}=\sqrt{Z} N$, $Z=1+\delta Z$, induces a four-nucleon term
\begin{eqnarray} \label{dL}
  \delta {\cal L} = - {C_0^{(s),{\rm finite}}}\,(2\delta Z)\, ( N^T P^{(s)}_i
N)^\dagger ( N^T P^{(s)}_i N)\,.
\end{eqnarray}
Since $\delta Z\sim Q_r^4\sim Q^2$ this term is treated
perturbatively.  A tree level counterterm
\begin{eqnarray}
 \delta^{\rm uv,0}D_2 = 9\, C_0^{\rm finite} {g_A^2 \over (4\pi
 f)^2} \Big[ \frac1\epsilon -\gamma_E +\ln(\pi) \Big]
\end{eqnarray}
is introduced to cancel the $1/\epsilon$ pole.  Dressing the
operator in Eq.~(\ref{dL}) with $C_0$ bubbles gives
\begin{eqnarray}  \label{wfnren}
  -9i \:[A^{(-1)}]^2\,{M\gamma \over 4\pi }\,{g_A^2 m_\pi^2 \over (4\pi
 f)^2}\, \bigg[\frac13 + \ln\Big({\mu^2 \over m_\pi^2}\Big)
 \bigg]\,.
\end{eqnarray}
Again, for $n=2$ we have $\delta Z=0$ so no new PDS counterterms were added. 
Note that if we had instead used bare nucleon fields then there would be no
correction of the form in Eq.~(\ref{dL}).  However, Eq.~(\ref{gph6}) would be
modified because the last graph in Fig.~\ref{Qr3}f is no longer present.  When
this is combined with the  contribution from the LSZ formula the sum of
Eq.~(\ref{gph6}) and Eq.~(\ref{wfnren}) is reproduced.

For PDS, the graphs in Figs.\ref{Qr3}a-f are finite for $n=2$ so no new finite
subtractions were introduced. For $n=3$, counterterms are introduced to
renormalize the terms with $\ln(\mu^2)$ in
Eqs.~(\ref{pt1},\ref{ans3},\ref{wfnren}) (in PDS $\mu=\mu_R$). In OS only terms
analytic in $m_\pi^2$ are subtracted \cite{ms1} (including $m_\pi^2
\ln(\mu^2)$).  We find $D_2(\mu_R)\to D_2(\mu_R)+\Delta D_2(\mu_R)$, with
\begin{eqnarray} \label{DD2}
  \Delta D_2(\mu_R)=  6\, C_0(\mu_R)^2\, {g_A^2 \over (4\pi f)^2}\,
    {M (\gamma-\gamma')\over 4\pi}\, \bigg[-\frac13 +\kappa+ \ln\Big({\mu_R^2 
    \over  \mu_0^2} \Big) \bigg]\,.
\end{eqnarray}
Here $\kappa=1/3$ in PDS and $\kappa=0$ in OS, and $\mu_0$ is an unknown
scale.  Note that the logarithm in Eq.~(\ref{DD2}) gives a contribution to the
beta function for $D_2(\mu_R)$ of the form
\begin{eqnarray}
  \beta_{D_2}^{(rad)} =
 { 3 g_A^2 \over  4\pi^2 f^2} \,  {M (\gamma-\gamma')\over 4\pi} \, C_0(\mu_R)^2\,.
\end{eqnarray}
This disagrees with the beta function of Ref.~\cite{ksw2}, because in that paper
the beta function was calculated including only the one-loop graphs.

Adding the contributions in Eqs.~(\ref{pt1},\ref{ans3},\ref{gph6},\ref{wfnren})
gives the total radiation pion contribution to the amplitude at order $Q_r^3$:
\begin{eqnarray} \label{fans}
i A^{rad} &=&6i\, [ A^{(-1)}]^2 {g_A^2 m_\pi^2 \over (4 \pi f)^2}\, 
  {M (\gamma-\gamma') \over 4\pi}\, \bigg[   \kappa + 
  {\rm ln}\Big({ {\mu_R}^2\over m_\pi^2 }\Big) \bigg]  
 -i \, [ A^{(-1)}]^2\: {\Delta D_2(\mu_R)\, m_\pi^2 \over C_0(\mu_R)^2} \nn \\ 
&&+i\, [ A^{(-1)}]^2 \bigg[{M(\gamma -\gamma')\over 4\pi}\bigg]^2 
  {g_A^2 \over \sqrt{\pi} f^2} \Big({m_\pi \over M}\Big)^{3/2} I_2 
  \Big( {E \over m_\pi} \Big)\,.
\end{eqnarray}
The first term here has the same dependence on the external momentum as an
insertion of the $D_2$ operator dressed by $C_0$ bubbles. Its $\mu_R$
dependence is cancelled by $\mu_R$ dependence in $\Delta D_2(\mu_R)$. Note
that due to cancellations between graphs, this term is actually suppressed by a
factor of $\gamma/Q_r$ relative to what one expects from the power counting. 
The second term in Eq.~(\ref{fans}) has a nontrivial dependence on $E$ and is
suppressed by an even smaller factor of $\gamma^2/Q_r^2$.  These cancellations
were not anticipated by the power counting and it would be interesting to
determine if the suppression by factors of $\gamma\sim1/a$ continues at higher
orders.  If so, this might be a consequence of an additional symmetry of the
theory in the limit $a\to \infty$.  If not, terms at order $Q_r^4$ may actually give 
the leading contribution of radiation pions to NN scattering.

If we now consider momenta $p\sim m_\pi \sim Q\ll Q_r$, we should fix $\mu_R$
at the threshold, $\mu_R=\sqrt{M m_\pi}$, and expand in $E/m_\pi$ giving
$I_2(E/m_\pi)=-3.94 +{\cal O}(E/m_\pi)$.  Therefore, the dominant effect of the
graphs that occur at order $Q_r^3$ is indistinguishable from a shift in
$D_2(\mu_R)$. Integrating out the radiation pions amounts to absorbing their
effects into the effective $D_2$ in the low energy theory.  The
result in Eq.~(\ref{fans}) is suppressed relative to the NLO contributions in
Ref.~\cite{ksw2} by a factor of roughly $2(\gamma-\gamma')/M\sim 1/10$.  Since
this is smaller than the expansion parameter, $Q/\Lambda\sim 1/3$, it can be
neglected at NNLO.  

Recall, that in evaluating the graphs in Fig.~\ref{Qr3} a mulitpole expansion is
used, which in this case is an expansion in $v=\sqrt{m_\pi/M}$.  If the first
correction in the multipole series were to multiple a nonzero term of order
$m_\pi^{1/2}$ then this would give an order $m_\pi$ contribution.  However, we
have checked that for all the graphs in Fig.~\ref{Qr3} the first correction in the
multipole  expansion gives a vanishing contribution.

\begin{figure}[!t]
  \centerline{\epsfysize=3.0truecm \epsfbox{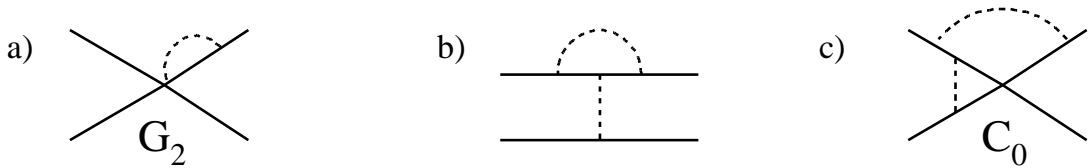}  }
 {\tighten
\caption[1]{Examples of order $Q_r^4$ radiation pion graphs for NN scattering.  
} \label{ho} }
\end{figure}
At order $Q_r^4$ graphs such as those in Fig.~\ref{ho} will contribute to NN 
scattering.  The graph in Fig.~\ref{ho}a includes an insertion of the operator
\begin{eqnarray}
{\cal L} = i\, G_2\, [ N^T P^{(s)}_i
N ]^\dagger\, [ N^T P^{(s)}_i  \sigma_j (\xi\partial_j\xi^\dagger -
\xi^\dagger\partial_j\xi) N] +h.c. \,.
\end{eqnarray}
(Note that due to the hermitian conjugate this operator is the same for $s=\!^1S_0$
and $s=\!^3S_1$.) This graph will be dressed with $C_0$ bubbles inside and outside
the radiation pion loop.  The renormalization group equation for $G_2$ gives
$G_2(\mu_R) \sim 1/(M\mu_R^2) \sim 1/(M Q_r^2)$.  Combining this with the
remaining factors of $Q_r$ we find that Fig.~\ref{ho}a is of order $Q_r^4/(M^4
\Lambda_\chi^2)$ and is therefore suppressed by $Q_r/M$ relative to a graph in
Fig.~\ref{Qr3}.  Power counting the graphs in Fig.~\ref{ho}b,c gives
$Q_r^4/(M^3\Lambda_{\rm NN} \Lambda_\chi^2)$, giving a factor of
$Q_r/\Lambda_{\rm NN}$ relative to a graph in Fig.~\ref{Qr3}.  This provides an
example of how graphs with potential pions seem to restrict the range of the
effective field theory to $ \Lambda_{\rm NN}\sim 300\,{\rm MeV}$.  The $300\,{\rm
MeV}$ scale applies only to a subset of graphs and may change once all graphs at
this order are included.

It is important to understand how a $Q_r^m$ correction scales with $Q$ for 
$p\sim m_\pi$.  In Ref.~\cite{msconf} it is demonstrated that $Q_r^m$ graphs
can give a $Q^{m/2-1}$ contribution plus terms higher order in $Q$.  For the
$Q_r^3$ calculations the leading terms scale as $Q^{1/2}$ and come from the 
graphs in Fig.~\ref{Qr3}b,d,e.  However, it turns out the these $Q^{1/2}$ terms
cancel.  It is not understood at present why this cancellation occurs.  Graphs
at order $Q_r^4$ may give an order $Q$ contribution for $p\sim m_\pi$ and 
should be included in a complete NNLO calculation of the NN phase shifts.

\subsection{Soft Pions}

In this section soft pion contributions will be discussed.  We will see that there are
graphs with non-vanishing soft contributions that should be included for $p
\gtrsim m_\pi$.  In soft loops two scales appear, $m_\pi$ and $p=M v$.  We will see
below that it is necessary to take $p\sim Q_r$ when power counting graphs with
soft loops in order to avoid double counting.  In other words $v$ should have the
same value as in the radiation pion calculation. A soft loop has energy and
momentum $q_0\sim q\sim Q_r$, so $d^4 q\sim Q_r^4$.  The mass of the soft pion
is smaller than its momentum, and is treated perturbatively.  Nucleon propagators
in a soft loop are static (like in heavy quark effective theory, see eg. 
Ref.~\cite{neubert}) since the loop energy is greater than the nucleon's kinetic
energy \cite{gries}.  Therefore, these propagators count as $1/Q_r$.  This power
counting is identical to that proposed in Ref.~\cite{ksw2} except powers of $Q_r$
are counted rather than $Q$.
 
Unlike potential pions, both soft and radiation pion pieces come from taking the
pole in a pion propagator.  Therefore, we must be careful not to double count when
adding these contributions.  This is accomplished by taking $p\sim Q_r$ when
evaluating both soft and radiation pion graphs.  This ensures that the soft and
radiation modes have different momenta ($\sim Q_r$ and $\sim m_\pi$
respectively).  Integrals involving the scales $Q_r$ and $m_\pi$ can be separated
using the method of asymptotic expansions and dimensional regularization
\cite{beneke,smirnov}.  Consider splitting a loop integral into two regimes by
introducing a momentum factorization scale $L$ such that $m_\pi < L < Q_r$.  After
the pion pole is taken in an energy integral over $q_0$, the remaining integral is of
the form 
\begin{eqnarray} \label{split}
  \int d^n q \ = \ \int_0^L d^n q \mbox{ (radiation)} + 
	\int_L^{\infty} d^nq \mbox{ (soft)} \,,
\end{eqnarray}
which is obviously independent of $L$.  In Eq.~(\ref{split}) the power counting
dictates that expansions in $m_\pi^2/Q_r^2$ should be made so that each integral
becomes a sum of integrals involving only one scale ($m_\pi$ for radiation and
$Q_r$ for soft).  In dimensional regularization power divergences vanish, while
logarithmic divergences show up as $1/\epsilon$ poles. Therefore, after expanding
we can take $L\to \infty$ in the radiation integral and $L\to 0$ in the soft integral. 
Taking the $L\to \infty$ and $L\to 0$ limits may introduce ultraviolet divergences
for the radiation integral and infrared divergences for the soft integral.   When we
add the radiation and soft contributions any superfluous $1/\epsilon$ poles will
cancel.  This will be illustrated with an explicit example below.  The asymptotic
expansion procedure has been rigorously proven for Feynman graphs with large
external Euclidean momenta and large masses \cite{smirnov2}.  It has also been
shown to work for the non-relativistic threshold expansion of one and two-loop
graphs \cite{beneke}.

Notice that it is crucial to expand the soft pion propagator in powers of
$m_\pi^2/Q_r^2$, because otherwise the radiation pion contribution may be double
counted.  As an example, consider the graph in Fig.~\ref{Qr3}a.  Taking $p\sim
Q_r$ implies $M v^2\sim m_\pi$.  For the radiation pion contribution $q_0
\sim q \sim M v^2 \sim m_\pi$, so we keep the $m_\pi^2$ in the denominator of
Eq.~(\ref{gr1}).  When computing the soft contribution, we assume $q_0\sim q\sim
Q_r \gg m_\pi$, and must expand the denominator in powers of $m_\pi^2/Q_r^2$. 
The $\vec q$ integration is now scaleless so the soft contribution to Fig.~\ref{Qr3}a
vanishes in dimensional regularization.  If we did not expand in $m_\pi/Q_r$ when
evaluating the soft contribution, we would have double counted the radiation
contribution.  The same argument can be applied to all the diagrams in
Fig.~\ref{Qr3}.  In each case the soft contribution vanishes. 

\begin{figure}[!t]
  \centerline{\epsfysize=3.0truecm \epsfbox{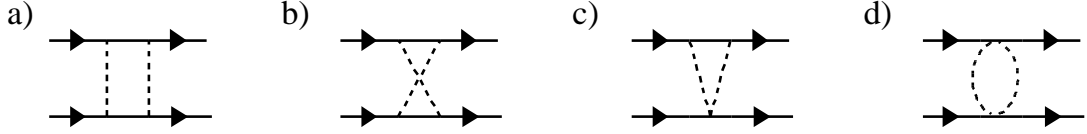}  }
 {\tighten
\caption[1]{Examples of one-loop graphs which have soft pion contributions. 
Graphs a)-d) also have a radiation pion contribution, while in addition
graph a) has a potential pion contribution. } \label{fig_soft} }
\end{figure}
Examples of graphs which have a non-vanishing soft contribution are shown in
Fig.~\ref{fig_soft}.  These diagrams were calculated in Ref.~\cite{kaiser} (although
the S-wave channels were not analyzed there).  In the KSW power counting they
must be dressed on the outside with $C_0$ bubbles.  (If Fig.~\ref{fig_soft}a or
\ref{fig_soft}b are dressed on the inside with $C_0$ bubbles then the soft
contribution vanishes.)  To see how these graphs obtain contributions from the
soft and radiation regimes consider Fig.~\ref{fig_soft}a.  Unlike the massless case
\cite{beneke}, this graph has a radiation contribution.  In the $^1S_0$ channel the 
loop integral for Fig.~\ref{fig_soft}a is
\begin{eqnarray}  \label{s1}
&& \bigg({-i g_A^2 \over 2 f^2}\bigg)^2 \int {d^D q \over
  (2 \pi)^D} {i \over \frac{E}2+q_0 -{({\vec q}+{\vec p})^2 \over 2M} +i
  \epsilon}\ {i \over \frac{E}2-q_0 -{({\vec q}+{\vec p})^2 \over 2M} +i
  \epsilon}\ { {\vec q\:}^2 \over q_0^2 - {\vec q}\:^2 - m_\pi^2 + i \epsilon}\nn\\ 
&& \qquad\qquad\qquad \times { ({\vec q-\vec t\,})^2 \over q_0^2 - 
   ({\vec q-\vec t\,})^2 - m_\pi^2 +  i \epsilon} \,,
\end{eqnarray}
where $\vec t = {\vec p\,}' -\vec p$ and $\pm \vec p$ and $\pm \vec p\,'$ are the
incoming and outgoing nucleon momenta.  Unlike the graphs in Fig.~\ref{Qr3}, we
are forced to route an external momentum, $\vec t$, through a pion propagator. 
Taking a nucleon pole in Eq.~(\ref{s1}) gives the potential pion contribution
proportional to $M p$.  Taking the contribution from the pion poles gives soft and
radiation contributions.  Our power counting tells us that the leading order soft
contribution will be $\sim Q_r^2$, while the leading order radiation contribution will
be $\sim Q_r^4/M^2$.  In Eq.~(\ref{s1}) the factors of $E/2-(\vec q+\vec p)^2/(2M)$ 
can be dropped.  In the soft regime the factors of $E/2-(\vec q+\vec p)^2/(2M)$ are
order $Q_r^2/M$, and are dropped relative to $q_0 \sim Q_r$ leaving static
nucleon propagators.  In the radiation regime $E/2-(\vec q+\vec p)^2/(2M)\to 0$ 
after using the multipole expansion and equations of motion.  This leaves
\begin{eqnarray}  \label{s2}
 && \frac{i}2 \bigg({g_A^2 \over 2 f^2}\bigg)^2 \int {d^n q \over
  (2 \pi)^n}\: {{\vec q}\:^2 (\vec q-\vec t\,)^2 \over 
 {\vec q}\,^2 - (\vec q -\vec t\,)^2}\: \Bigg\{ {1 \over 
 [{\vec q}\:^2 + m_\pi^2]^{3/2} } -{1\over [ ({\vec q -\vec t\,})^2 + m_\pi^2]^{3/2} }
  \Bigg\} \nn\\[5pt]
 &&=  \frac{i}2 \bigg({g_A^2 \over 2 f^2}\bigg)^2 \int {d^n q \over
  (2 \pi)^n}\: {{\vec q}\:^2 (\vec q-\vec t\,)^2 \over [{\vec q}\:^2 + m_\pi^2]^{3/2} }
  \bigg\{ {1\over {\vec q}\:^2 - (\vec q -\vec t\,)^2 + i\epsilon } + 
  {1\over {\vec q}\:^2 - (\vec q -\vec t\,)^2 - i\epsilon } \bigg\} \,,
\end{eqnarray}
where $n=D-1$.  The singularity at ${\vec q}\:^2 = (\vec q -\vec t\,)^2$ is cancelled
only in the sum of terms in the first line of Eq.~(\ref{s2}).  These terms can be
calculated separately by introducing an $i\epsilon$ in this denominator
\cite{beneke}, giving an average over $\pm i\epsilon$ as indicated\footnote{\tighten
The second line of Eq.~(\ref{s2}) is more easily split into soft and radiation
contributions.  If we had used the integrand on the first line we would also have to
consider $(\vec q-\vec t)\,^2\sim m_\pi^2$.}.  The factor of $(\vec q-\vec t\,)^2$ in
the numerator can be removed by partial fractioning.  For the soft contribution the
scale of the loop momentum is set by the external momentum, $q_0 \sim |\vec q|
\sim |\vec t| \sim Q_r$. Expanding Eq.~(\ref{s2}) in $m_\pi^2/{\vec q}\:^2$ gives
\begin{eqnarray}  \label{boxs}
  && i \bigg({g_A^2 \over 2 f^2}\bigg)^2 \int {d^n q \over (2 \pi)^n}\: { |\vec q| \over
  (2 \vec q \cdot \vec t - {\vec t\,}^2 \pm i\epsilon) } \sum_{m=0}^\infty  
  {\Gamma(-1/2)  \over \Gamma(-1/2-m) \Gamma(m+1)} \bigg( \frac{m_\pi^2}
  {{\vec q\:}^2} \bigg)^m  \nn\\[5pt] 
  && = {-i \over 192 \pi^2} \bigg({g_A^2 \over 2 f^2}\bigg)^2 \Bigg\{ \bigg[ 
  \frac{1}{\epsilon} + \ln\bigg(\frac{\bar \mu^2}{{t}^2}\bigg) \bigg] \Big(t^2 
  -18\, m_\pi^2\Big) - \bigg[ 
  \frac{1}{\epsilon} + \ln\bigg(\frac{\bar \mu^2}{{t}^2}\bigg) \bigg] \bigg( 90\, 
  {m_\pi^4 \over t^2} - 140\, {m_\pi^6 \over t^4} +\ldots \bigg) \nn\\*
  && \qquad\qquad\qquad\qquad  + \frac83\, t^2 -36\, m_\pi^2 +280\, 
      \frac{m_\pi^6}{t^4} + \ldots \Bigg\} \,,
\end{eqnarray}
where we have kept the first few terms in the expansion.  The soft
contribution starts at order $Q_r^2$ as expected.  The first $1/\epsilon$ pole in
Eq.~(\ref{boxs}) is an ultraviolet divergence, while the second is an infrared
divergence.  For the radiation contribution $q_0 \sim |\vec q| \sim m_\pi \ll |\vec t|$. 
Expanding in $(2\vec t \cdot \vec q) / {\vec t}\:^2$ gives
\begin{eqnarray}  \label{boxr}
  && -i \bigg({g_A^2 \over 2 f^2}\bigg)^2 \int {d^n q \over (2 \pi)^n}\: { 1 \over
  [{\vec q}\:^2 + m_\pi^2]^{3/2}  } \Bigg[\ {\vec q\:}^2 + { {\vec q\:}^4 \over 
  {\vec t\:}^2 } \sum_{m=0}^\infty \bigg( \frac{2 \vec q \cdot \vec t } 
  {{\vec t\:}^2\pm i\epsilon} \bigg)^m \ \Bigg] \nn\\[5pt] 
  && = {-i \over 192 \pi^2} \bigg({g_A^2 \over 2 f^2}\bigg)^2 \Bigg\{ -72\, m_\pi^2
  \bigg[ \frac{1}{\epsilon} + \ln\bigg(\frac{\bar \mu^2}{m_\pi^2}\bigg) \bigg] 
  + \bigg[  \frac{1}{\epsilon} + \ln\bigg(\frac{\bar \mu^2}{m_\pi^2}\bigg) \bigg] 
  \bigg( 90\, {m_\pi^4 \over t^2} - 140\, {m_\pi^6 \over t^4} +\ldots \bigg)\nn\\*
  && \qquad\qquad\qquad\qquad  -24\, m_\pi^2 +39\, \frac{m_\pi^4}{t^2} -
  \frac{482}{3} \, \frac{m_\pi^6}{t^4} + \ldots \Bigg\} \,.
\end{eqnarray}
The radiation contribution starts out as order $Q_r^4/M^2$ as expected.  Note that
only powers of $m_\pi=Q_r^2/M$ appear.  The $1/\epsilon$ poles in Eq.~(\ref{boxr})
are ultraviolet divergences.  When the soft and radiation contributions are added
the infrared poles in Eq.~(\ref{boxs}) cancel a subset of the ultraviolet poles in
Eq.~(\ref{boxr}).  Adding Eq.~(\ref{boxs}) and Eq.~(\ref{boxr}) we find
\begin{eqnarray}  \label{boxsr}
  &&  {-i \over 192 \pi^2} \bigg({g_A^2 \over 2 f^2}\bigg)^2 \Bigg\{ t^2 \bigg[ 
  \frac{1}{\epsilon} + \ln\bigg(\frac{\bar \mu^2}{{t}^2}\bigg) \bigg]+\frac83\,t^2 
  -m_\pi^2  \bigg[ \frac{90}{\epsilon} +18 \ln\bigg(\frac{\bar \mu^2}{{t}^2}\bigg)
  +72 \ln\bigg(\frac{\bar \mu^2}{{m_\pi}^2}\bigg) \bigg] -60\, m_\pi^2 \nn\\
 && \qquad\qquad\qquad\qquad +\ln\Big(\frac{t^2}{m_\pi^2}\Big) \bigg( 90\, 
  {m_\pi^4 \over t^2} - 140\, {m_\pi^6 \over t^4} +\ldots \bigg) +39\, 
      \frac{m_\pi^4}{t^2} +\frac{358}{3} \,  \frac{m_\pi^6}{t^4} + \ldots \Bigg\} \,,
\end{eqnarray}
where the remaining ultraviolet $1/ \epsilon$ poles are cancelled by counterterms
for $C_2$ and $D_2$.

If we are interested in making predictions for $p\sim m_\pi$, then powers of 
$m_\pi^2/t^2$ must be summed up.  Summing the series in Eq.~(\ref{boxsr}) gives
\begin{eqnarray}  \label{boxsum}
  &3a)=&  {-i \over 192 \pi^2} \bigg({g_A^2 \over 2 f^2}\bigg)^2 \Bigg\{ \Big(t^2
  -90\,m_\pi^2\Big) \bigg[ \frac{1}{\epsilon} + \ln\bigg(\frac{\bar \mu^2}{{m_\pi}^2}
  \bigg) \bigg]+\frac83\,t^2 -58 m_\pi^2 \nn\\  
&& \qquad\qquad\qquad\qquad - {(128 m_\pi^4 +16 m_\pi^2 t^2 -t^4)\over
   t \sqrt{t^2 + 4m_\pi^2} } \ln\bigg({\sqrt{t^2 + 4m_\pi^2}-t \over 
   \sqrt{t^2 + 4m_\pi^2} +t }\bigg) \Bigg\} \,. 
\end{eqnarray}
Since the coefficients in the series in Eqs.~(\ref{boxs},\ref{boxr}) diverge for $\vec
p\:' = \vec p$, Eq.~(\ref{boxsum}) should be used when integrating over $dt$ to
obtain the $^1S_0$ partial wave amplitude (even for $p\sim Q_r$). 
Eq.~(\ref{boxsum}) agrees with the result of evaluating Eq.~(\ref{s2}) exactly. 
Although the asymptotic expansion is not necessary for evaluating
Fig.~\ref{fig_soft}a, it allows us to identify the radiation and soft contributions to
this graph and verify that the power counting for each regime works.  We also see
that adding soft and radiation pions reproduces the correct answer without double
counting.  Recall that power counting with $p\sim Q_r$ was necessary to avoid
double counting for graphs like those in Fig.~\ref{Qr3}.  

For $p\sim Q_r$ the diagrams in Fig.~\ref{fig_soft} are order $Q_r^2/(f^2
\Lambda_\chi^2)$, and are larger than the order $Q_r^3$ graphs with a single
radiation pion in Fig.~\ref{Qr3}.  Decreasing $p$ to $p\sim m_\pi$ the graphs in
Fig.~\ref{fig_soft} give contributions of the form
\begin{eqnarray}
  {  {\vec t\,}^2 \over f^2 \, (4\pi f)^2  } \
	F\Big({{\vec t\:}^2 / m_\pi^2}\Big) \,,
\end{eqnarray}
where $F$ is a function.  For $p\sim m_\pi$ the graphs in Fig.~\ref{fig_soft} are
order $m_\pi^2$ which is smaller than the graphs in Fig.~\ref{Qr3} which
include order $m_\pi^{1/2}$, $m_\pi$, and $m_\pi^{3/2}$ terms.  It is interesting
to note that the relative importance of these graphs changes with $p$.  The graphs
in Fig.~\ref{fig_soft} dressed by $C_0$ bubbles give a contribution that is the same
size as a four nucleon operator with 6 derivatives, $C_6(\mu_R) p^6$, and are
N$^3$LO in the KSW power counting.   

It would be nice to see the expansion used in evaluating the radiation contribution
to Fig.~\ref{fig_soft}a implemented at the level of the Lagrangian.  It is not clear to
us how to do this at the present time.  In the radiation regime, the pion whose pole
is taken can be thought of as a radiation pion.  However, the other propagator
gives factors of $1/t^2$, ${\vec q}\cdot {\vec t} /t^4$, etc., which look more like
insertions of non-local operators than the propagator of a field.  Also, since in
general $\vec p \ne \vec p\:'$, the couplings for this second propagator change the
nucleon momenta and therefore do not involve a multipole expansion.  Finally, the
result in Eq.~(\ref{boxsum}) does not have an expansion in $E/m_\pi$.  So unlike
the radiation pion contribution computed in section A, this contribution cannot be
integrated out for $p<\sqrt{M m_\pi}$.   For these reasons, the use of the term
radiation for this contribution differs somewhat from the usage in section A.

To summarize, we have introduced a power counting in factors of $Q_r=\sqrt{M
m_\pi}$ appropriate for graphs with radiation pions.  The order $Q_r^3$ radiation
contributions to NN scattering were computed and found to be suppressed by
inverse powers of the scattering length.  Soft pion contributions also have a power
counting in $Q_r$.  For $p\sim Q_r$ they are $\sim Q_r^2$, but are higher order
than the radiation contributions for $p\sim m_\pi$.  Higher order corrections are
suppressed by factors of $Q_r/\Lambda$ and whether or not this expansion is
convergent is an open question.  If the range of the two-nucleon effective field
theory with perturbative pions is really $300\,{\rm MeV}$, and the suppression by
factors of $\gamma$ does not persist at higher orders then contributions from
radiation pions induce an incalculable error of order $m_\pi^2/\Lambda_\chi^2$ to
the NN scattering amplitude in this theory.  The validity of the $Q_r/\Lambda$
expansion can be tested by looking at processes at $p\sim 300\,{\rm MeV}$ such as
those with external pions.

We would like to thank Mike Luke and Mark Wise for several enlightening
discussions, as well as S.  Fleming, D.B. Kaplan and U. van Kolck for their
comments.  We also thank H. Griesshammer for emphasizing the importance of soft
pions.  This work was supported in part by the Department of Energy under grant
number DE-FG03-92-ER 40701.  T.M. was also supported by a John A.  McCone
Fellowship.

{\tighten

} 

\end{document}